# Quantum solid phase and Coulomb drag in two-dimensional electron-electron bilayers of MoS$_2$

*Meizhen Huang$^{1,4}$, Zefei Wu$^{1,4\,*}$, Ning Wang$^{1,3*}$ and Siu-Tat Chui$^{2*}$*

$^1$Department of Physics, Hong Kong University of Science and Technology, Clear Water Bay, Hong Kong

$^2$Bartol Research Institute and Department of Physics and Astronomy, University of Delaware, Newark, DE 19716, USA

$^3$William Mong Institute of Nano Science and Technology, the Hong Kong University of Science and Technology, Clear Water Bay, Hong Kong

$^4$These authors contributed equally: Meizhen Huang, Zefei Wu

$^*$Corresponding authors: phwang@ust.hk (NW), chui@udel.edu (STC) and zefei.wu@manchester.ac.uk (ZW)

## Abstract

**Coulomb drag experiments can give us information about the interaction state of double-layer systems. Here, we demonstrate anomalous Coulomb drag behaviours in a two-dimensional electron-electron bilayer system constructed by stacking atomically thin MoS$_2$ on opposite sides of thin dielectric layers of boron nitride. In the low temperature (*T*) regime, the measured drag resistance does not follow the behaviour predicted by the Coulomb drag models of exchanging momenta and energies with the particles in Fermi-liquid bilayer systems. Instead, it shows an upturn to higher and higher values. We investigate quantum solid/fluid phases and the Kosterlitz-Thouless/Wigner two-dimensional quantum melting transition in this bilayer system and describe this interesting phenomenon based on thermally activated carriers of quantum defects from the formation of the correlation-induced electron solid phases with enhanced stabilization by the potential due to the boron nitride dielectric layers.**

# 1. Introduction

Coulomb drag experiments can give us information about the effect of the long-range Coulomb interaction between charge particles in two closely spaced but electrically isolated low-dimensional electronic systems.[1] The transport characteristics of Coulomb drag are generally reflected by the drag resistance $R_{drag}$, defined as the ratio of the drag voltage $V_{drag}$ (in the passive layer) to the drive current $I_{drive}$ (in the active layer) as illustrated in **Figure 1**a. Coulomb drag measurements can directly show the interactions of charged particles both between the two coupled electronic systems and inside them,[1] and have become part of the standard toolbox in condensed matter physics widely used to investigate various properties in low dimensional electronic systems such as quantum metal-insulator transition,[2-5] exciton effects[6-12] and quantum coherence.[13, 14]

In two-dimensional Fermi liquid (FL) systems[15-18] (e.g., semiconductor heterostructures of GaAs double-quantum wells and graphene double layers), the dominant drag mechanism is the momentum transfer between carriers in the two layers by mutual Coulomb scattering: $R_{drag}$ is usually proportional to the electron-electron momentum relaxation time[16] and should vanish as $T^2$ when the temperature $T$ goes to zero. However, in semiconductor bilayers with larger interlayer spacing $d$, only very small momentum can be transferred. In this condition, as much as 30% of the measured $R_{drag}$ was ascribed to phonon-mediated interactions, where both energy and momentum transfer takes place.[1] In graphene bilayers (separated by an atomically thin boron nitride (BN) dielectric layer), long-range Coulomb interaction between charge particles is largely enhanced and the friction drag dominates over the phonon drag, resulting in giant momentum transfer between the two layers.[13, 18, 19] However, due to the high Fermi velocity and small effective mass, the Bohr radius $a_B = 90$ Å for the graphene system is comparable to that in GaAs. The dimensionless parameter for distance, the Wigner Seitz radius $r_s$ that is measured relative to $a_B$, is extremely small for experimentally accessible electron densities in graphene.[20] The physics is dominated by the quantum kinetic energy that is proportional to $1/r_s^2$. It is therefore difficult to study the physics of the effect of the Coulomb interaction which is proportional to $1/r_s$ and becomes important at low densities.

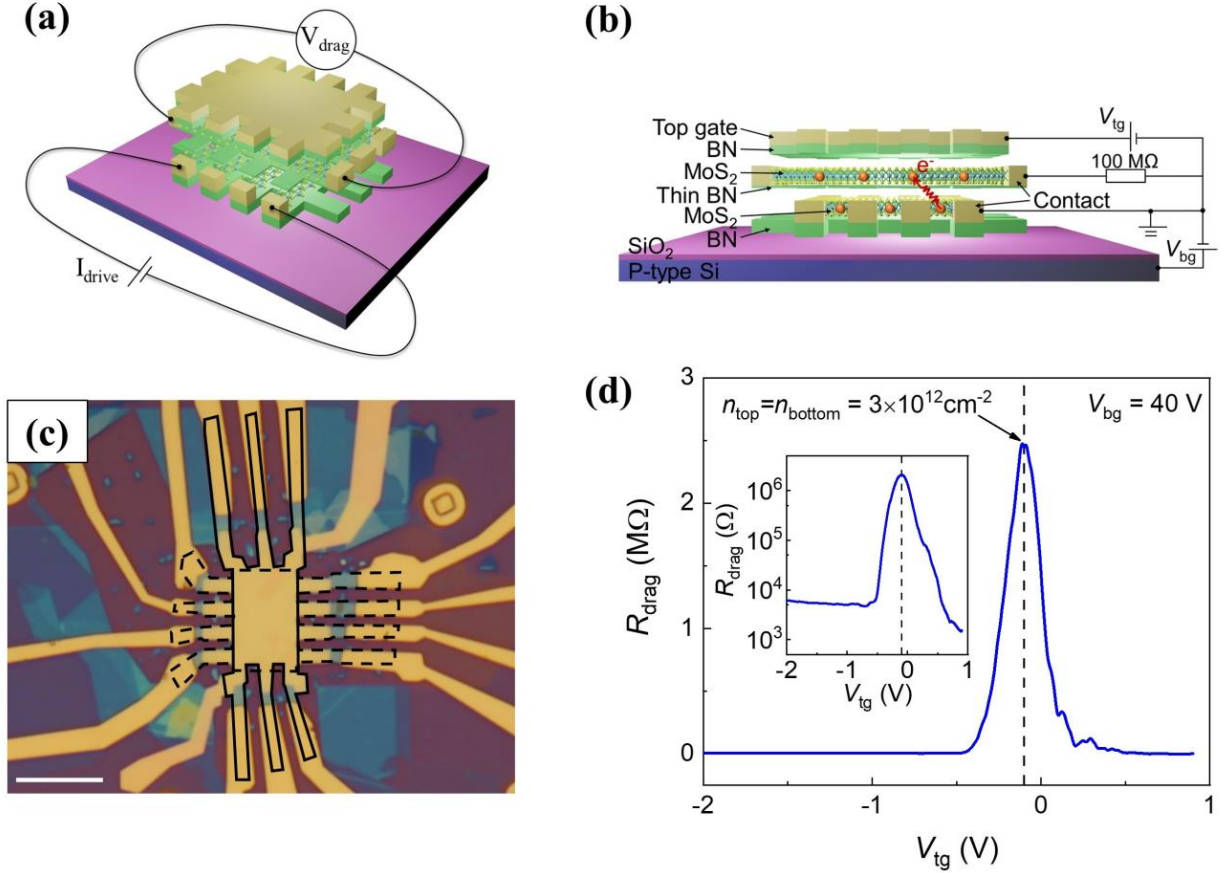

**Figure 1.** Coulomb drag in MoS$_2$ bilayer. a) The schematic device structure of the electron-electron bilayer system with current injection in the bottom layer and drag voltage measured from the top layer. b) Cross-section of the device showing atomically thin MoS$_2$ layers separated by an insulating barrier (BN) and encapsulated by the top and bottom cladding layer. c) The optical image of the device. The solid (dashed) line represents the top (bottom) layer, and the scale bar is 10 μm. d) Drag resistance measured at different top gate voltages at 1.5 K. Inset: Log plot of drag resistance. The drag resistance is ~kΩ when the carrier density for the two layers mismatch.

Transition metal dichalcogenide (TMDC) semiconductors have geometries similar to that of graphene and possess many valuable properties (e.g., large effective masses and strong electron-electron interactions with $a_B = 6$ Å for the experimental TMDC systems). TMDCs process a $r_s$ of the order of 10 for the experimentally accessible electron densities ($\sim 10^{12}/\text{cm}^2$),[21] one order of magnitude larger than that in monolayer graphene, thus opening up new dimensionless density regimes to explore Coulomb drag properties experimentally. In this work, we demonstrate anomalous drag behaviours in MoS$_2$/BN/MoS$_2$ electron-electron bilayers. At cryogenic temperatures, the measured drag resistance $R_{drag}$ neither vanishes nor follows the $T^2$ behaviour. Instead, it shows an upturn to higher and higher values. We attribute the unexpected Coulomb drag behaviour to the formation of bilayer quantum solid phases of electrons and their possible stabilization by the potential due to the BN dielectric material.

## 2. Experimental Section

Atomically thin bilayer devices, as shown in Figure 1b and 1c, were constructed through the mechanical exfoliation and dry transferring method developed previously.[22] Few-layer MoS$_2$ flakes (4L ~ 8L) were exfoliated onto SiO$_2$/Si substrates, separated by a thin potential barrier layer of BN (~5nm thick) and sandwiched by thicker BN (10~20nm thick) layers. Transport measurements were performed in a cryogenic system which provided stable temperatures ranging from 1.5 to 300 K. The top and bottom gates, $V_{tg}$ and $V_{bg}$, controlled the carrier densities in the two layers (dual-gate tunable layer resistance can be found in Figure S1, we have set the chemical potential of the fluid/solid phases to zero as estimated from the gate voltages and the corresponding electron densities). An electrical current $I_{drive}$ was injected into the bottom active layer, and the Coulomb drag voltages $V_{drag}$ were measured from the top passive layer. The $V_{drag}$-$I_{drive}$ characteristic (Figure S2) and the interlayer leakage current (Figure S3) are checked to confirm the validity of the drag measurements. We first measure the drag resistance $R_{drag}$ versus $V_{tg}$ (or equivalently the carrier density in the layer) at a fixed $V_{bg}$. Figure 1d shows a typical $R_{drag}$ obtained at 1.5K. Strikingly, $R_{drag}$ reaches a value of ~2.5MΩ at the 1:1 ratio of the carrier densities $n_{\text{top}} = n_{\text{bottom}} = n = 3 \times 10^{12} \text{cm}^{-2}$, which is three orders of magnitude larger than ~1kΩ when the carrier densities of two layers mismatch (as shown in inset of Figure 1d). One possible explanation for this $R_{drag}$ peak is the formation of the bilayer quantum solid phase. Previous studies show that when the Coulomb interaction between electrons dominates their kinetic energy, electrons can condense into a close-packed lattice at a low enough temperature and result in a quantum solid phase.[23-25] In our case, when the layer densities are about 1:1 matched, electrons in two layers can form solids that are commensurate and most effectively and completely coupled. The charge carriers are frozen out and resulting in an extremely high $R_{drag}$. The carrier density and temperature dependence of the drag signal are then performed to study the details of the drag behaviour.

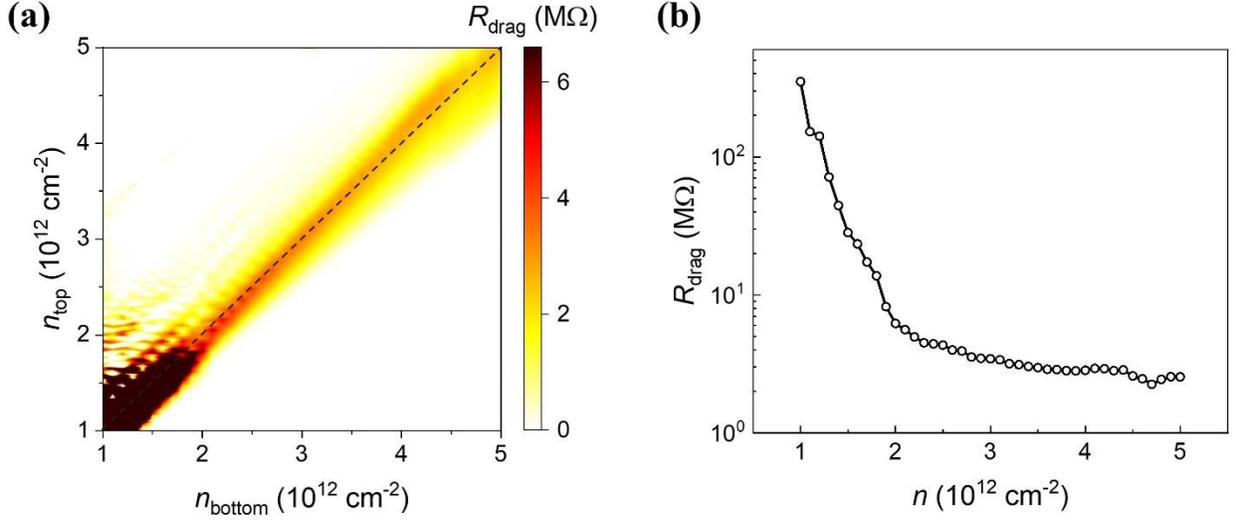

**Figure 2.** Carrier density dependence of the drag resistance. a) $R_{drag}$ plotted as a function of the carrier densities in the top (passive layer) and bottom layers. b) The amplitudes of drag resistance peaks at different carrier density $n = n_{\text{top}} = n_{\text{bottom}}$, the lower the matched density the higher the drag resistance peak.

$R_{drag}$ at other top and bottom carrier densities are measured and shown in **Figure 2**a, where the density matched peaks occur in the experimentally accessible range of carrier density of 1-5×$10^{12}$/cm$^2$. The amplitudes of $R_{drag}$ peaks are carrier density dependent, the lower the matched density the higher the $R_{drag}$ (Figure 2b). At the low carrier density region (1-2×$10^{12}$/cm$^2$), $R_{drag}$ increases exponentially with decreasing carrier density. We explained this based on the picture of thermally activated quantum defects in the quantum solid state. In the solid state, the free electrons are frozen, and the carriers can come from thermally activated quantum defects.[26] The effect of such defects was first emphasized by Kosterlitz and Thouless, who proposed that in two dimensions the solid phase can be characterized by a finite shear modulus, even though there is no long-range order.[27] The zero shear modulus in the fluid phase occurs when dislocation pairs in the solid phase become unbound. A different idea of the stabilization of the solid came from Wigner, who pointed out that the potential energy gained is more than the quantum kinetic energy lost for the solid at low densities.[28] We found that defects such as dislocations in electron solids are quantum and form waves with an effective mass of the order of the electron mass.[26] The energy of a dislocation in the electron solid can be lowered by coupling to the zero point motion of the phonons.[29] This lowering becomes big at low densities, thus combining the Wigner idea and the Kosterlitz Thouless picture. An example of a defect responsible for the transport is a dislocation pair,[26, 29] with an energy approximately given by $\Delta \cong \frac{e^2}{1.9\,\epsilon a_B r_s}\left[0.048\left(1-\frac{r_{s0}}{r_s}\right)+0.021\right]$ for a constant $r_{s0}$. The exponential increase of $R_{drag}$ can be a result of the term of $\Delta$ that is proportional to $1/r_s^2$; as the density $n$ is decreased, the solid defect energy is strengthened due to the decrease of the quantum kinetic energy from the zero point motion. The carrier density decreases and $R_{drag}$ is increased.

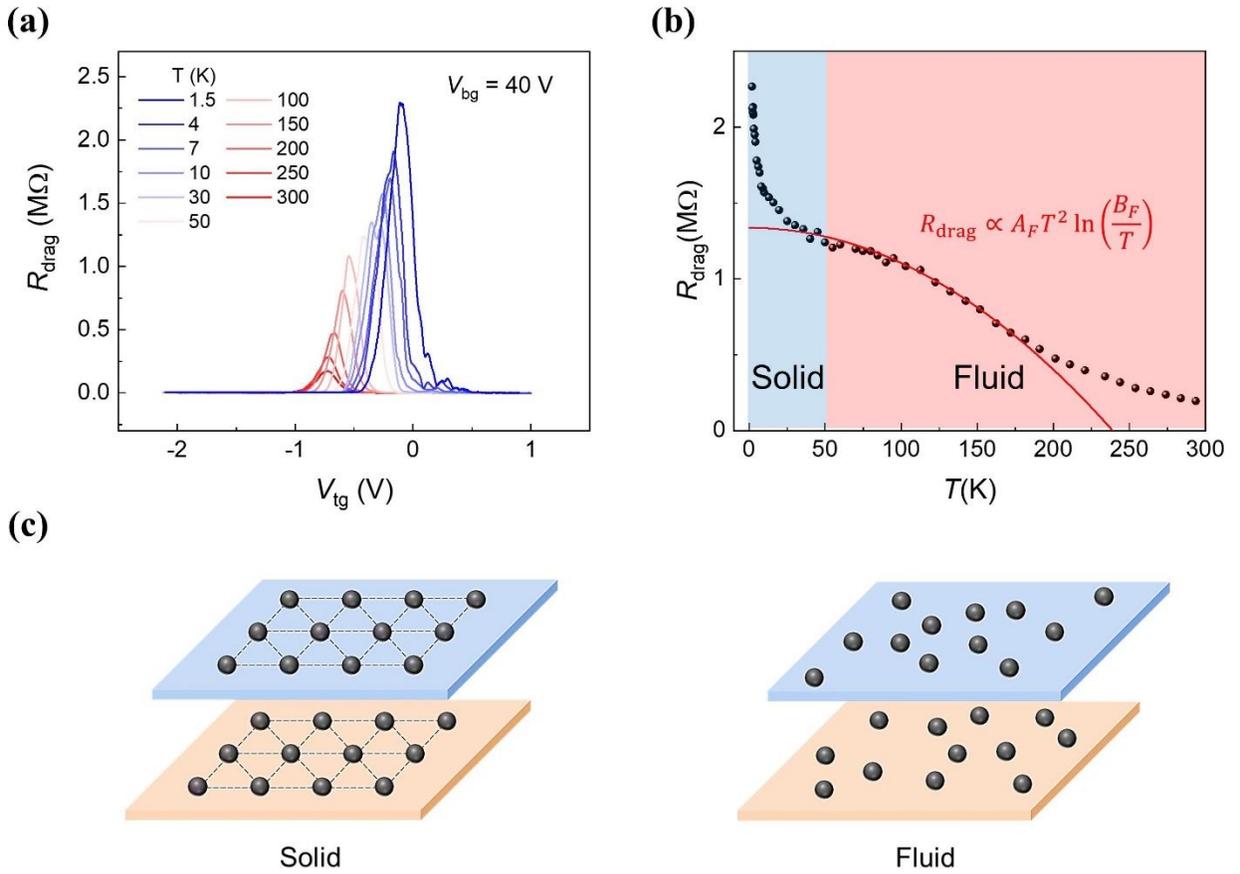

**Figure 3.** Transition between quantum solid and quantum fluid. a) Temperature dependence of the drag resistance. b) Temperature dependence of the drag peak amplitudes. Circles are experimentally measured values and the red line is the fit using the diffusive model ($A_F = 0.3$, $B_F = 1.5 \times 10^{-32}$). c) Schematic diagrams of quantum solid and quantum fluid phase, with localized and itinerant electrons, respectively. Black circles represent electrons inside the top (blue) and bottom (orange) layers

**Figure 3** shows the temperature-dependent $R_{drag}$, where the amplitudes of the $R_{drag}$ peaks continuously increase with decreasing sample temperature. First, a striking increase at ultra-low temperatures (1.5~50K) is observed. Obviously, $R_{drag}$ at this regime is anomalous. It does not follow the $T^2$ dependence in the ballistic model or the $T^2 \ln\frac{1}{T}$ dependence in the diffusive model, implying a new Coulomb drag mechanism different from the momentum transferring model of FL bilayer systems. Since the interlayer electron-electron interaction dominates their kinetic energy at low enough temperature, electrons can form a bilayer quantum solid phase (as shown in Figure 3c). We explained this anomalous $R_{drag}$ drag on the formation of the correlation-induced solid phases as shown below. Second, when the temperature is higher than the melting temperature (50~175K), the system transits into a quantum fluid phase. In this regime, our experimental results can be well fitted by the diffusive model,[1] demonstrating a robust drag signal from momentum transfer between the MoS$_2$ layers. Further increasing the temperature (175~300K), the approximation that the temperature is much less than the Fermi temperature becomes invalid. Thermally excited quasiparticles and plasmons coexist and give additional contributions to the drag signal, thus the observed $R_{drag}$ is slightly higher than the value predicted by the diffusive model.

## 3. Discussion and Modelling

The anomalous upturn of the drag resistance has been observed in electron-hole bilayer systems such as GaAs-AlGaAs[6] and graphene-GaAs[30] where electron-hole pairing and possible exciton condensation play important roles. However, the mechanism of this anomalous behavior in our system should be totally different from theirs, as electron-hole pairing cannot happen in our electron-electron system. We attribute the anomalous Coulomb drag behavior at low temperatures to the formation of MoS$_2$ two-dimensional bilayer crystal phases of electrons. The low-temperature dependence in Figure 3b can be understood as follows. As we explained, we expect the transport to be carried by thermally activated defects of a density that is proportional to $e^{-\Delta/kT}$. This results in a $R_{drag} \sim e^{\Delta/kT}$. As shown in **Figure 4**, our experimental results can be well fitted by this formula, demonstrating the formation of the inter-layer solid phase in our sample. To improve our understanding, we have performed numerical studies of our system with no adjustable parameters as described below.

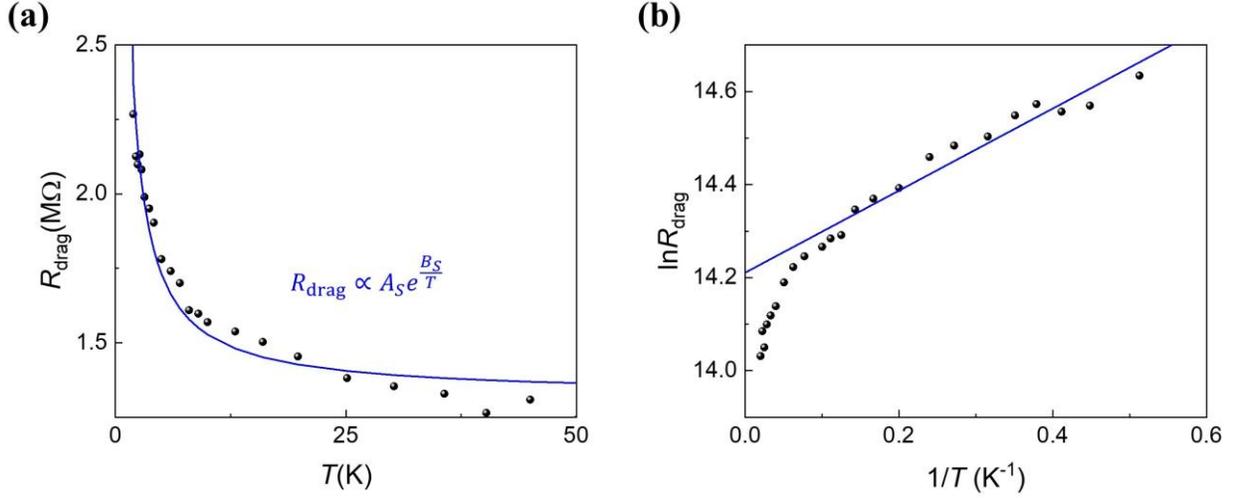

**Figure 4.** $R_{\text{drag}}$ in low-temperature regime plotted using a different scale. Circles are experimental results. The blue line in (a) is a fitting using $A_S = 4.9 \times 10^8 \Omega$, $B_S = 0.88$K, in (b) a linear fitting.

We have calculated the quantum solid and fluid energies and their transition in this bilayer system with fixed node diffusion Monte Carlo method,[31, 32] taking into account the realistic experimental structure including the screening effect[33-35] of the BN spacer, the encapsulation and the substrate.[36] As mentioned previously, the smaller Bohr radius for the TMDC systems favors the formation of the crystal electronic phases at a much higher electron density. Previous numerical[37-40] and analytical studies[32] showed that the interlayer Coulomb interaction enhances the formation of the Wigner crystal phases and extends the maximum solidification density. Those numerical studies were motivated by the study of GaAs heterostructures and did not correspond to the experimentally available parameter range of the TMDC systems. However, compared to the formation of intra-layer Wigner crystals, a coupled bilayer system has an enhanced Coulomb potential energy which is favorable to the formation of inter-layer crystal phases. Here, we consider one more favourable factor, the BN spacer with thickness $d$ stacked between the two $MoS_2$ layers which is critical for the formation of the coupled bilayer device structure.

The energies of the solid and the fluid phases for a system with 30 particles on each side of the structure for different densities and BN spacer thicknesses $d$ are calculated. In **Figure 5**a, we show the difference between the solid and the fluid energies as a function of $r_s$ that is defined in terms of the density of each layer for $d = 5$nm. We have also carried out calculations with the effect of the BN potential by considering static Coulomb potential from the periodic array of boron and nitrogen ions (details of the calculation process can be found in our previous paper[41]). The energy difference between the solid and the fluid phases is shown in Figure 5b. From where the energy difference is equal to zero, we obtain the phase boundary of the quantum melting transition of the present system and the critical Wigner Seitz radius $r_{sc}$. The critical $r_{sc}$ for three different cases are shown in Figure 5c. As expected, the bilayer system tends to stabilize the Wigner solid phase, resulting in a transition at a dimensionless distance $r_{sc} \approx 8.4$, much smaller than the value of 30 for the single-layer TMDC case. In addition, the consideration of the BN potential results in a $r_{sc} \approx 7.4$, the smallest value in these cases. The stabilization of the Wigner solid phase here is partly due to the interlayer coupling effect.

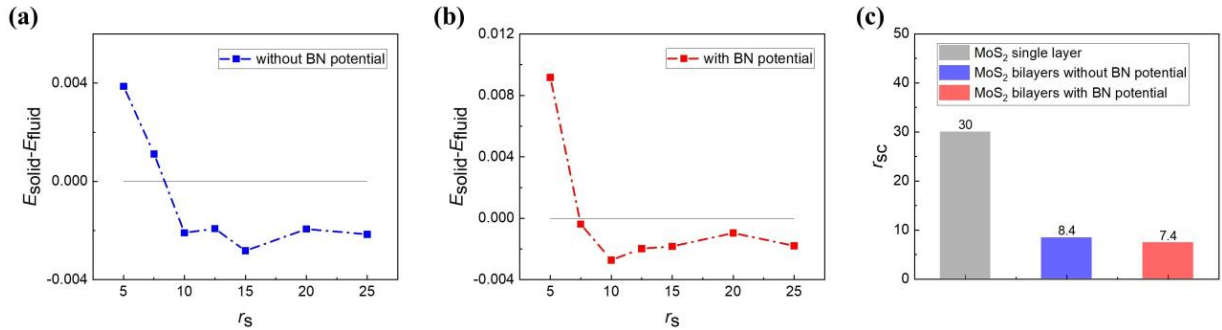

**Figure 5.** Calculation of the difference between the solid and the fluid energies as a function of $r_s$ for $d = 5$nm (a) without and (b) with BN potential. c) The solid-fluid transition $r_{sc}$ for three different cases.

## 4. Conclusion

In conclusion, we demonstrate that the Coulomb drag resistance measured from atomically thin MoS$_2$ bilayers shows an upturn when decreasing the temperature, very different from the Coulomb drag models of exchanging momenta and energies with the particles in Fermi-liquid bilayer systems. We describe the anomalous Coulomb drag behaviours based on the properties of the quantum defects of the electron solid and their possible stabilization by the potential caused by the BN dielectric layers.

**Supporting Information**
Supporting Information is available from the Wiley Online Library or from the author.

**References**

[1]  B.N. Narozhny, A. Levchenko, *Rev. Mod. Phys.* **2016**, 88, 025003.
[2]  C. Jörger, S.J. Cheng, W. Dietsche, R. Gerhardts, P. Specht, K. Eberl, K. von Klitzing, *Phys. E* **2000**, 6, 598.
[3]  C. Jörger, S.J. Cheng, H. Rubel, W. Dietsche, R. Gerhardts, P. Specht, K. Eberl, K.v. Klitzing, *Phys. Rev. B.* **2000**, 62, 1572.
[4]  R. Pillarisetty, H. Noh, E. Tutuc, E.P. De Poortere, K. Lai, D.C. Tsui, M. Shayegan, *Phys. Rev. B.* **2005**, 71, 115307.
[5]  R. Tao, L. Li, H.-Y. Xie, X. Fan, L. Guo, L. Zhu, Y. Yan, Z. Zhang, C. Zeng, *arXiv e-prints* **2020**, arXiv:2003.12826.
[6]  A.F. Croxall, K. Das Gupta, C.A. Nicoll, M. Thangaraj, H.E. Beere, I. Farrer, D.A. Ritchie, M. Pepper, *Phys. Rev. Lett.* **2008**, 101, 246801.
[7]  J.A. Seamons, C.P. Morath, J.L. Reno, M.P. Lilly, *Phys. Rev. Lett.* **2009**, 102, 026804.
[8]  J.A. Seamons, D.R. Tibbetts, J.L. Reno, M.P. Lilly, *Appl. Phys. Lett.* **2007**, 90, 052103.
[9]  E. Tutuc, M. Shayegan, D.A. Huse, *Phys. Rev. Lett.* **2004**, 93, 036802.
[10] J.P. Eisenstein, A.D.K. Finck, D. Nandi, L.N. Pfeiffer, K.W. West, *J. Phys.: Conf. Ser.* **2013**, 456, 012009.
[11] J.I.A. Li, T. Taniguchi, K. Watanabe, J. Hone, C.R. Dean, *Nat. Phys.* **2017**, 13, 751.
[12] C. Zhang, G. Jin, *J. Phys.: Condens. Matter* **2013**, 25, 425604.
[13] S. Kim, I. Jo, J. Nah, Z. Yao, S.K. Banerjee, E. Tutuc, *Phys. Rev. B.* **2011**, 83, 161401.
[14] A.S. Price, A.K. Savchenko, B.N. Narozhny, G. Allison, D.A. Ritchie, *Science* **2007**, 316, 99.
[15] J.P. Eisenstein, *Superlattices Microstruct.* **1992**, 12, 107.
[16] T.J. Gramila, J.P. Eisenstein, A.H. MacDonald, L.N. Pfeiffer, K.W. West, *Phys. Rev. Lett.* **1991**, 66, 1216.
[17] T.J. Gramila, J.P. Eisenstein, A.H. MacDonald, L.N. Pfeiffer, K.W. West, *Phys. B* **1994**, 197, 442.
[18] L. Zhu, L. Li, R. Tao, X. Fan, X. Wan, C. Zeng, *Nano Lett.* **2020**, 20, 1396.
[19] K. Lee, J. Xue, D.C. Dillen, K. Watanabe, T. Taniguchi, E. Tutuc, *Phys. Rev. Lett.* **2016**, 117, 046803.
[20] S. Das Sarma, S. Adam, E.H. Hwang, E. Rossi, *Rev. Mod. Phys.* **2011**, 83, 407.
[21] J. Lin, T. Han, B.A. Piot, Z. Wu, S. Xu, G. Long, L. An, P. Cheung, P.-P. Zheng, P. Plochocka, X. Dai, D.K. Maude, F. Zhang, N. Wang, *Nano Lett.* **2019**, 19, 1736.
[22] Z. Wu, B.T. Zhou, X. Cai, P. Cheung, G.-B. Liu, M. Huang, J. Lin, T. Han, L. An, Y. Wang, S. Xu, G. Long, C. Cheng, K.T. Law, F. Zhang, N. Wang, *Nat. Commun.* **2019**, 10, 611.
[23] T. Smoleński, P.E. Dolgirev, C. Kuhlenkamp, A. Popert, Y. Shimazaki, P. Back, X. Lu, M. Kroner, K. Watanabe, T. Taniguchi, I. Esterlis, E. Demler, A. Imamoğlu, *Nature* **2021**, 595, 53.

[24] E.C. Regan, D. Wang, C. Jin, M.I. Bakti Utama, B. Gao, X. Wei, S. Zhao, W. Zhao, Z. Zhang, K. Yumigeta, M. Blei, J.D. Carlström, K. Watanabe, T. Taniguchi, S. Tongay, M. Crommie, A. Zettl, F. Wang, *Nature* **2020**, 579, 359.
[25] M. Kumar, A. Laitinen, P. Hakonen, *Nat. Commun.* **2018**, 9, 2776.
[26] S.T. Chui, K. Esfarjani, *Phys. Rev. Lett.* **1991**, 66, 652.
[27] J.M. Kosterlitz, D.J. Thouless, *J. Phys. C: Solid State Phys.* **1973**, 6, 1181.
[28] E. Wigner, *Phys. Rev.* **1934**, 46, 1002.
[29] S.T. Chui, K. Esfarjani, *Europhys. Lett.* **1991**, 14, 361.
[30] A. Gamucci, D. Spirito, M. Carrega, B. Karmakar, A. Lombardo, M. Bruna, L.N. Pfeiffer, K.W. West, A.C. Ferrari, M. Polini, V. Pellegrini, *Nat. Commun.* **2014**, 5, 5824.
[31] D. Ceperley, *Phys. Rev. B.* **1978**, 18, 3126.
[32] S.T. Chui, B. Tanatar, *Phys. Rev. Lett.* **1995**, 74, 458.
[33] L.V. Keldysh, *Jetp Lett.* **1979**, 29, 658.
[34] A. Chernikov, T.C. Berkelbach, H.M. Hill, A. Rigosi, Y. Li, O.B. Aslan, D.R. Reichman, M.S. Hybertsen, T.F. Heinz, *Phys. Rev. Lett.* **2014**, 113, 076802.
[35] C. Xiao, F. Wang, S.A. Yang, Y. Lu, Y. Feng, S. Zhang, *Adv. Funct. Mater.* **2018**, 28, 1707383.
[36] Y. Lu, W. Xu, M. Zeng, G. Yao, L. Shen, M. Yang, Z. Luo, F. Pan, K. Wu, T. Das, P. He, J. Jiang, J. Martin, Y.P. Feng, H. Lin, X.-s. Wang, *Nano Lett.* **2015**, 15, 80.
[37] S. De Palo, F. Rapisarda, G. Senatore, *Phys. Rev. Lett.* **2002**, 88, 206401.
[38] G. Senatore, F. Rapisarda, S. Conti, *Int. J. Mod. Phys. B* **1999**, 13, 479.
[39] F. Rapisarda, G. Senatore, *Aust. J. Phys.* **1996**, 49, 161.
[40] S. Narasimhan, T.-L. Ho, *Phys. Rev. B.* **1995**, 52, 12291.
[41] S.T. Chui, N. Wang, C.Y. Wan, *Phys. Rev. B.* **2020**, 102, 125420.